# High Proton Conductivity of $H_xWO_3$ at Intermediate Temperatures: Unlocking Its Application as a Mixed Ionic–Electronic Conductor

*Rantaro Matsuo[1], Tomoyuki Yamasaki[1]\*, and Takahisa Omata[1]\**

Institute of Multidisciplinary Research for Advanced Materials (IMRAM),
Tohoku University, 2-1-1 Katahira, Aoba-ku, Sendai 980-8577, Japan.

ABSTRACT. Hydrogen tungsten bronzes ($H_xWO_3$), known for their mixed protonic–electronic conduction near room temperature, are extensively studied for electrochromic and gasochromic applications. However, their proton transport properties at elevated temperatures—particularly in the intermediate-temperature range (200–500 °C)—remain unexplored. This study revealed the proton transport behavior of $H_xWO_3$, focusing on its potential as a proton-conducting mixed ionic–electronic conductor (MIEC) for intermediate-temperature electrochemical applications. By employing a proton-conducting phosphate glass as an electron-blocking electrode, we selectively measured the partial proton conductivity of sintered $H_xWO_3$. Hydrogen incorporation into the sintered $WO_3$ pellet was found to occur preferentially near the surface, forming an approximately 500-µm-thick hydrogen-rich region. This region reached a composition of $x = 0.24$ and exhibited proton conductivity exceeding $10^{-1}$ S cm$^{-1}$ at 275 °C—well above those of the state-of-the-art perovskite proton conductors. Impedance spectroscopy revealed distinct features of proton transport, including an isotope effect. The proton diffusion coefficient was 100–1000 times greater than that of $H_{\sim 0.0001}TiO_2$, which exhibits mixed protonic–electronic conduction via hydrogen dissolution. The larger proton diffusion coefficient of $H_{0.24}WO_3$ suggests that large polaron formation enhances proton mobility. These findings unlock new functionality of $H_xWO_3$ as a MIEC in the intermediate-temperature range, paving the way for the development of next-generation hydrogen energy conversion systems.

INTRODUCTION

Tungsten trioxide ($WO_3$) is a well-known transition metal oxide that exhibits pronounced optical and electronic changes upon hydrogen incorporation[1–10]. When exposed to hydrogen gas at elevated temperatures or subjected to electrochemical hydrogen injection using acidic electrolytes, $WO_3$ forms hydrogen tungsten bronzes ($H_xWO_3$). $WO_3$ can accommodate a large amount of hydrogen (up to $x \sim 0.5$)[11], corresponding to proton density ($n_H$) on the order of $10^{21}$ cm$^{-3}$. Upon hydrogen dissolution, hydrogen atoms are incorporated into the lattice and ionize into protons and electrons, as described by Equation (1) in Kröger–Vink notation[12]:

$$1/2\, H_2 \rightarrow H_i^\times \rightarrow H_i^\cdot + e' \quad (1)$$

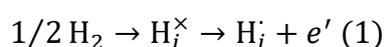

The electrons introduced along with protons localize on tungsten atoms, leading to the reduction of $W^{6+}$ ions to $W^{5+}$. These changes result in a characteristic blue coloration and an increase in electronic conductivity. These coupled optical and electronic changes have been widely utilized in electrochromic[13–15] and gasochromic devices[16,17], hydrogen sensors[18–20], and resistive switching devices that operate via ion injection[21,22].

Since the optical and electronic responses in $H_xWO_3$ are associated with proton diffusion processes, $H_xWO_3$ is characterized as a proton-conducting mixed ionic–electronic conductor (MIEC), and the dynamics of proton transport have attracted significant research interest. The proton diffusion process has been studied using various techniques such as proton nuclear magnetic resonance ($^1$H NMR)[23–27] and electrochemical measurements[28–30]. Changes in optical absorption[31–33] have been also utilized as experimentally accessible indicators of proton migration. These previous studies have revealed that the proton diffusion coefficient in $H_xWO_3$ strongly depends on crystallinity, hydrogen content, and measurement techniques and have consistently reported that proton diffusion in $H_xWO_3$ is significantly faster than that in many other oxides[34–36]. Despite these attractive properties, the practical applications of $H_xWO_3$ are limited to gasochromic and electrochromic devices operating near room temperature. Only a few reports have explored other uses of its mixed conduction[37,38].

Owing to its intrinsic mixed protonic–electronic conductivity, $H_xWO_3$ is a promising candidate for electrochemical devices operating at elevated temperatures, such as protonic ceramic fuel cells and steam electrolysis systems[39]. In such devices, mixed protonic–electronic conductors are particularly advantageous because they allow electrochemical reactions to extend beyond narrow triple-phase boundaries (gas–electrolyte–electrode) to broader two-phase boundaries (gas–electrode)[40]. This can reduce electrode polarization resistance and improve the overall efficiency, which is especially beneficial for operation in the intermediate-temperature range (200–500 °C), where sluggish electrode kinetics often limit performance. Thus, demonstrating high proton conductivity of $H_xWO_3$ in the intermediate-temperature range would unlock the potential use of $H_xWO_3$ as a MIEC, contributing to the development of next-generation energy devices such as hydrogen permeable membranes and fuel cells.

This study investigates the proton conductivity of $H_xWO_3$ in the intermediate-temperature range using an electron-blocking method. A proton-conducting phosphate glass with a proton transport number of unity[41] is employed as an electron-blocking electrode, enabling selective detection of protonic conduction in $H_xWO_3$ while suppressing electronic contributions. The partial proton conductivity of sintered $H_xWO_3$ is above $10^{-2}$ S cm$^{-1}$ at 250 °C, which significantly exceeds that of well-known perovskite-type proton conducting oxides. The large proton diffusion coefficient responsible for this high proton conductivity is discussed in terms of the thermodynamics of diffusion, particularly the contributions of activation enthalpy and activation entropy.

EXPERIMENTAL

**Sample Preparation**

The sintered $WO_3$ samples were prepared by spark plasma sintering (SPS; SPS-511S, Fuji Electronic Industrial Co., Ltd., Japan). The raw yellowish green $WO_3$ powder (99.9%, Kojundo Chemical Laboratory Co., Ltd., Japan) was ground using a mortar and pestle made of stabilized $ZrO_2$, put into a carbon mold with a diameter of 20 mm, and then sintered at 800 °C under 50 MPa for 10 min in vacuum. All sintered pellets were polished to a uniform thickness of 2.3 mm. The as-sintered sample had a black appearance, owing to oxygen deficiency caused by annealing under an atmosphere with low oxygen chemical potential. After subsequent annealing in pure oxygen at 600 °C for 40 h, the sinters turned back yellowish green, indicating that oxygen vacancies were filled with oxygen. The apparent density of the sinters, determined by the Archimedes method, was approximately 99% of the theoretical density. To promote hydrogen incorporation into the sintered $WO_3$ samples, a 100-nm-thick Pd film was deposited on both faces by magnetron sputtering (JFC-1600, JEOL, Japan), followed by annealing in a hydrogen atmosphere at 300 °C for up to 144 h. After hydrogen annealing, the surface of the sinter turned dark blue, indicating the formation of $W^{5+}$ due to hydrogen incorporation, i.e., the formation of $H_xWO_3$. A cross-sectional scanning electron microscopy image (JSM-7800F, JEOL, Japan) of $H_xWO_3$ revealed a dense structure with no visible cracks or voids (Figure S1). Grain boundaries were too fine to be observed, suggesting a dense and homogeneous microstructure.

**Evaluation of Hydrogen Incorporation into Sintered $WO_3$**

The crystalline phase in the sintered $H_xWO_3$ was identified by X-ray diffraction (XRD; SmartLab, Rigaku, Japan) using Cu-Kα radiation. Prior to measurement, the sintered pellet was mechanically polished to obtain a flat surface for accurate diffraction analysis. To evaluate the depth profile of hydrogen, a two-dimensional hydrogen concentration map was acquired by using time-of-flight secondary ion mass spectrometry (TOF-SIMS; TOF-SIMS5, ION-TOF GmbH, Germany) over a 500-μm-wide region from the outer surface of the cross-sectional face of the $H_xWO_3$ pellet. The amount of hydrogen dissolved in the sintered $H_xWO_3$ was quantified by thermal desorption spectrometry (TDS) using a quadrupole mass spectrometer (QMASS; M-200QA, ANELVA, Japan). A small piece of the sintered sample (approximately 0.1 g), either oxygen- or hydrogen-annealed, was placed in a $SiO_2$-glass crucible, which was then set in a closed-end alumina tube. The tube was placed in a furnace and connected to an analyzing chamber equipped with a QMASS. After the experimental apparatus, comprising the sample and analyzing chambers, was evacuated by using a turbomolecular pump (TMU-261, Pfeiffer Vacuum, Germany) down to <1×10$^{-7}$ Pa at room temperature, the furnace was heated up to 1000 °C at a rate of 2.5 °C min$^{-1}$. H$_2$O ($m/z$ = 18) and H$_2$ ($m/z$ = 2) gases released from the samples were detected by the QMASS and quantified using La(OH)$_3$ and TiH$_2$ as standard samples. Details of the TDS apparatus and experimental procedure have been described elsewhere[42].

**Partial Proton Conductivity Measurements**

The partial proton conductivity was measured by the electron-blocking method using the proton-conducting phosphate glass (36HO$_{1/2}$−4NbO$_{5/2}$−2BaO−4LaO$_{3/2}$−4GeO$_2$−1BO$_{3/2}$−49PO$_{5/2}$ glass) with a thickness of 20–30 µm and a proton transport number of unity[41] as the electron-blocking electrode. The junction between the sintered H$_x$WO$_3$ and the glass was fabricated through hot-pressing, as previously reported[43,44]. To reduce interfacial resistance across the junction between the sintered H$_x$WO$_3$ and the glass, a 100-nm-thick Pd buffer layer deposited by rf-magnetron sputtering (1″ sputter cathode, Kenix, Japan) was inserted in the junction. Next, 100-nm-thick Pd reversible electrodes for protons and electrons were deposited on both ends of the cell by rf-magnetron sputtering. The conductivity was measured by using both DC and AC-impedance methods (VersaSTAT-3F, Princeton Applied Research, USA) at 200−275 °C under H$_2$ or deuterium (D$_2$) atmosphere. The applied voltage for the DC measurement was 100 mV. For AC-impedance measurements, the frequency range varied from 100 mHz to 1 MHz with an amplitude of 10 mV.

RESULTS AND DISCUSSION

**Hydrogen Incorporation into Sintered WO$_3$**

Figure 1(a) presents the XRD patterns obtained from the surface and core regions of the WO$_3$ pellet annealed in H$_2$ for 144 h. The inset shows a cross-sectional image of the pellet, with the measured regions corresponding to the XRD data. The cross-section reveals a sharp transition from the dark blue surface to a pale blue core. The diffraction peaks from the surface and core are indexed to the tetragonal and monoclinic phases of WO$_3$, respectively[45,46]. The blue coloration deepens as the hydrogen content increases[4], and its crystal structure varies accordingly—monoclinic for $x \lesssim 0.1$, orthorhombic for $0.1 \lesssim x \lesssim 0.15$, and tetragonal for $0.15 \lesssim x \lesssim 0.5$[1,47]. These results suggest that the outer surface of the pellet is hydrogen-rich, while the inner core is hydrogen-poor.

Figure 1(b) displays the hydrogen depth profile of the sinter annealed in H$_2$ for 144 h, obtained by lateral averaging a two-dimensional hydrogen distribution map measured by using TOF-SIMS. The hydrogen signal is less detectable at the edge of the pellet, but a hydrogen-rich region is clearly observed near the surface. The concentration gradually begins to decrease around a depth of 500 µm, corresponding to the color boundary, and reaches a lower constant level toward the core. The hydrogen concentration is nearly constant within both the hydrogen-rich surface and the hydrogen-poor core, with no apparent internal gradients observed in the respective regions. Further details on the TOF-SIMS measurement are provided in Figure S2.

The hydrogen-rich surface and hydrogen-poor core regions were separated through mechanical grinding, and the hydrogen content in each region was determined by TDS. The hydrogen content was determined to be $x$ = 0.24 in the hydrogen-rich surface and $x$ = 0.0048 in the hydrogen-poor core (Figure S3). The presence of two regions with distinct hydrogen concentrations, separated by an intermediate region exhibiting a concentration gradient in the depth direction, suggests that the proton diffusion coefficient varies across regions with different hydrogen concentrations. According to Fick's law (Equation (2)), steeper concentration gradients correspond to lower local diffusion coefficients:

$$J = -D(c)\frac{dc(x)}{dx} \quad (2)$$

where $J$ is the proton flux, $D(c)$ is the proton diffusion coefficient as a function of the local proton concentration $c$, and $dc(x)/dx$ represents the hydrogen concentration gradient along the depth coordinate $x$. Similar behavior has also been reported in studies on the gasochromic properties of H$_x$WO$_3$ thin films, with a drop in the diffusion coefficient at the monoclinic/orthorhombic–tetragonal phase boundary[33]. The present study results and previous observations regarding $H_xWO_3$ thin films suggest that structural changes induced by hydrogen incorporation suppress the inward extension of the hydrogen-rich region. As shown in Figure S4, the hydrogen-rich region gradually extended inward with increasing annealing time. Even after 144 h of hydrogen annealing, the hydrogen-rich region extended only to approximately 500 μm from the pellet surface. Although the detailed mechanism of this behavior is beyond the scope of this study, it is worth noting that, at the phase boundary, proton diffusion must be accompanied by the tilt of $WO_6$ octahedra necessary for the monoclinic/orthorhombic-to-tetragonal phase transition[46]. Such coupling between proton migration and lattice deformation may result in reduced proton mobility, which explains the lower diffusion coefficient in the phase boundary[48].

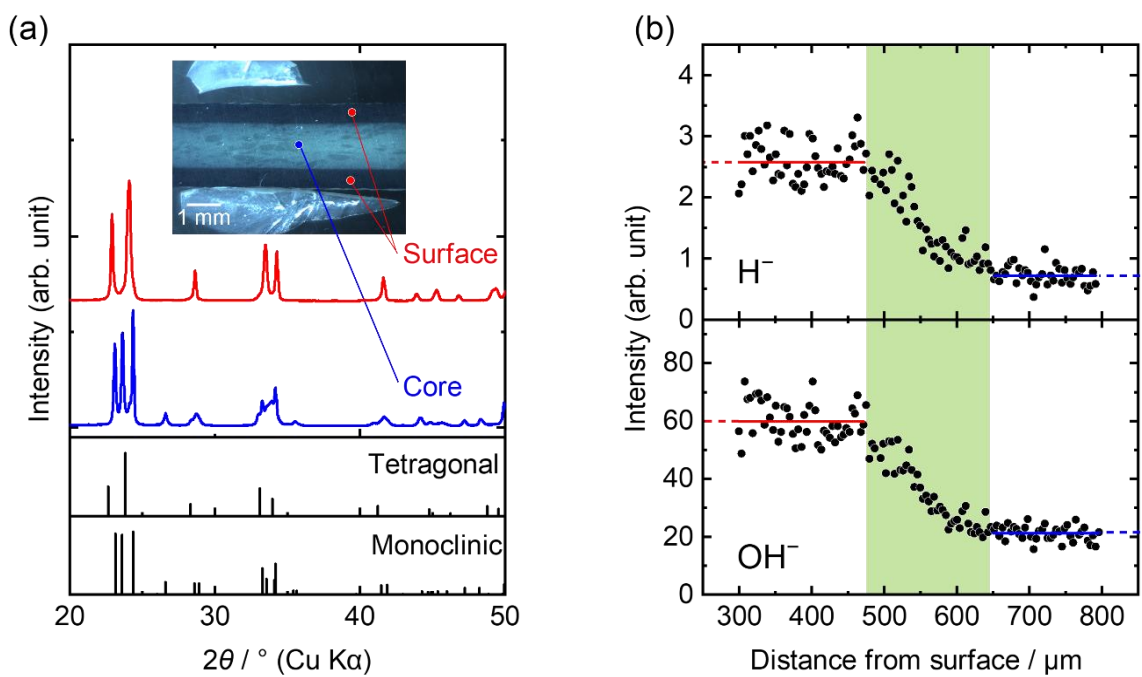

**Figure 1.** (a) XRD patterns of the surface (red) and core regions (blue) of the $H_xWO_3$ pellet. The inset shows a cross-sectional optical microscope image of the pellet, revealing a sharp color transition from the dark blue surface to the pale blue core. (b) Depth profile of hydrogen-related ions (H⁻ and OH⁻) obtained by TOF-SIMS near the phase boundary in the cross section of $H_xWO_3$.

**Proton Conductivity of H$_x$WO$_3$ in Hydrogen Atmosphere**

Figure 2(a) shows a schematic illustration of the electron-blocking cell, together with a top-view photograph of the sample. The phosphate glass was tightly laminated onto the surface of the sintered H$_x$WO$_3$, forming a well-defined interface that allows reliable characterization of proton conduction under electron-blocking conditions (Figure S5). Figure 2(b) shows the current–time profile measured under the application of a constant DC voltage of 100 mV as the atmosphere was switched between H$_2$ and D$_2$. The increase and decrease in current upon switching between H$_2$ and D$_2$ clearly demonstrate an isotope effect, suggesting that protons contribute to the current and that electronic conduction is effectively suppressed by the phosphate glass.

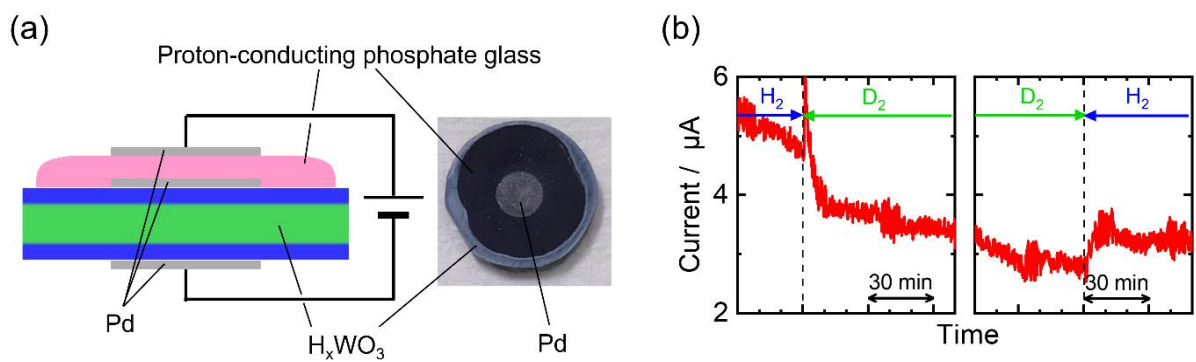

**Figure 2.** (a) Schematic image of the electron-blocking cell along with a top-view photograph. (b) Time evolution of the DC current (under a 100-mV applied bias) during the sequential switching of the atmosphere from H$_2$ to D$_2$ and back to H$_2$ at 250 °C.

Figure 3(a) shows the decay of the transient current immediately after the application of a DC voltage of 100 mV, indicating that the electron-blocking cell contains not only a resistive component associated with proton transport, but also a capacitive component. To examine the details of this behavior, the frequency-dependent responses of the cell were measured by AC impedance spectroscopy. Figure 3(b) shows the overall Nyquist plot of the electron-blocking cell, revealing the following distinct components: a high-frequency intercept resistance ($R_1$), mid-frequency arcs, and a low-frequency inclined straight-line response. Among these, the straight line observed in the low-frequency region has a slope close to 1, indicative of Warburg impedance ($R_w$)[49]. As shown in the enlarged view in Figure 3(c), two distinct arcs are observed in the mid-frequency region. These are denoted as $R_2$ and $R_3$ and are visually highlighted in red and blue, respectively, for clarity.

The equivalent circuit of the asymmetric electron-blocking cell can be described by a complex transmission line model that represents mass and charge transport, as shown in Figure S6[50]. The model includes elements related to charge transport at interfacial regions, leading to a complex impedance response. However, this complex impedance response makes it difficult to readily identify the resistive component associated with proton transport in $H_xWO_3$. Although the transmission line model provides a physically meaningful representation of the system, previous studies have shown that fitting with a series of Randles circuits yields nearly equivalent results[44,51]. Therefore, while the transmission line model is more appropriate for physical interpretation, we herein discuss the fitting results based on the simplified equivalent circuit shown in the upper part of Figure 3(c).

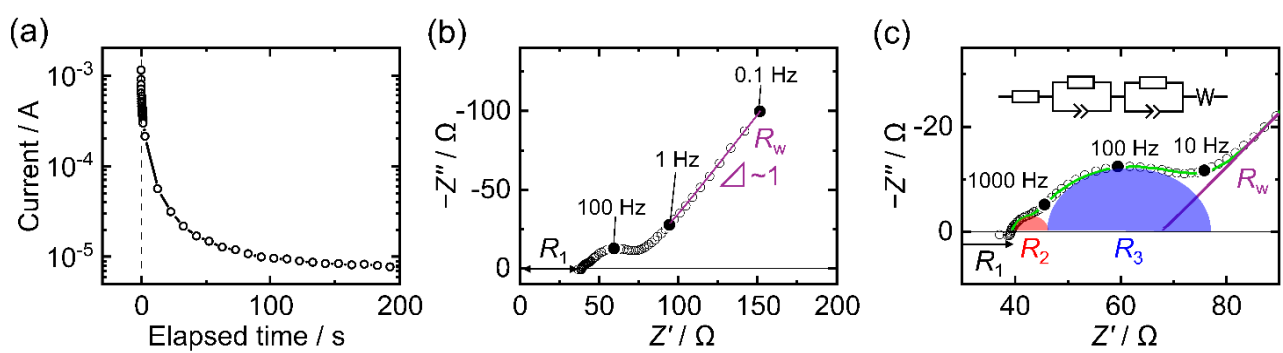

**Figure 3.** (a) Current decay after applying a 100-mV DC bias under a hydrogen atmosphere at 250 °C. (b) Nyquist plot of the AC impedance measured under the same conditions as in (a). (c) Enlarged view of the mid-frequency region in (b) along with the simplified equivalent circuit (top) used for fitting and the fitting result (green line).

As a first step, the contribution from proton transport in the glass electrode can be readily separated, as its relaxation is the fastest and appears as a distinct high-frequency resistance. As shown in Figure S7, the temperature dependence of $R_1$ agrees well with the reported value of the proton-conducting glass[41], allowing it to be assigned to proton transport within the glass. The remaining impedance components were identified through the subsequent measurements.

Figure 4(a) presents the Nyquist plots of the electron-blocking cell measured under $H_2$ and $D_2$ atmospheres. Switching the atmosphere from $H_2$ to $D_2$ increased not only $R_1$, reflecting the isotope effect in the glass electrode, but also $R_2$ and $R_3$. This phenomenon provides evidence that these arcs are associated with proton-related processes such as proton transport in $H_xWO_3$ or interfacial charge transfer. The corresponding resistance values are plotted in Figure 4(b).

To further elucidate the assignment of the $R_2$ and $R_3$ components, we performed impedance measurements on $H_xWO_3$ pellets of different thicknesses. The pellet with an initial thickness of 2.3 mm was mechanically polished by 500 µm from the side opposite to the glass layer to reduce its thickness to 1.8 mm. The corresponding Nyquist plots are shown in Figure 4(c), and the changes in the $R_1$, $R_2$, and $R_3$ components are summarized in Figure 4(d). As noted in Figure S3, the thickness of the hydrogen-rich region is approximately 500 µm. Therefore, removing an area of approximately 500 µm from one side of the pellet effectively reduces the thickness of the hydrogen-rich region by about half, while leaving the hydrogen-poor core nearly unchanged. This reduction primarily affected the $R_2$ component, which suggests that $R_2$ corresponds to the proton transport resistance within the hydrogen-rich region.

In contrast, the $R_3$ component exhibited only a minor change upon thickness reduction, suggesting a possible contribution from proton transport in the hydrogen-poor core, the thickness of which remained nearly unchanged. However, since the polishing was performed from the side opposite to the glass electrode, the glass-side interface remained unchanged. Thus, contributions to $R_3$ from charge transfer resistance at the Pd/glass and Pd/$H_x$WO$_3$ interfaces cannot be ruled out. Therefore, the present results are insufficient to unambiguously identify the processes responsible for $R_3$ and determine the proton conductivity of the hydrogen-poor core.

Finally, $R_w$ exhibited negligible change upon isotope substitution or variation in pellet thickness, suggesting the presence of diffusion-limited proton transport within the cell, regardless of sample geometry or atmosphere. This result is in line with the presence of a diffusion-retarding region at the monoclinic–tetragonal phase boundary, as described in Figure 1(b).

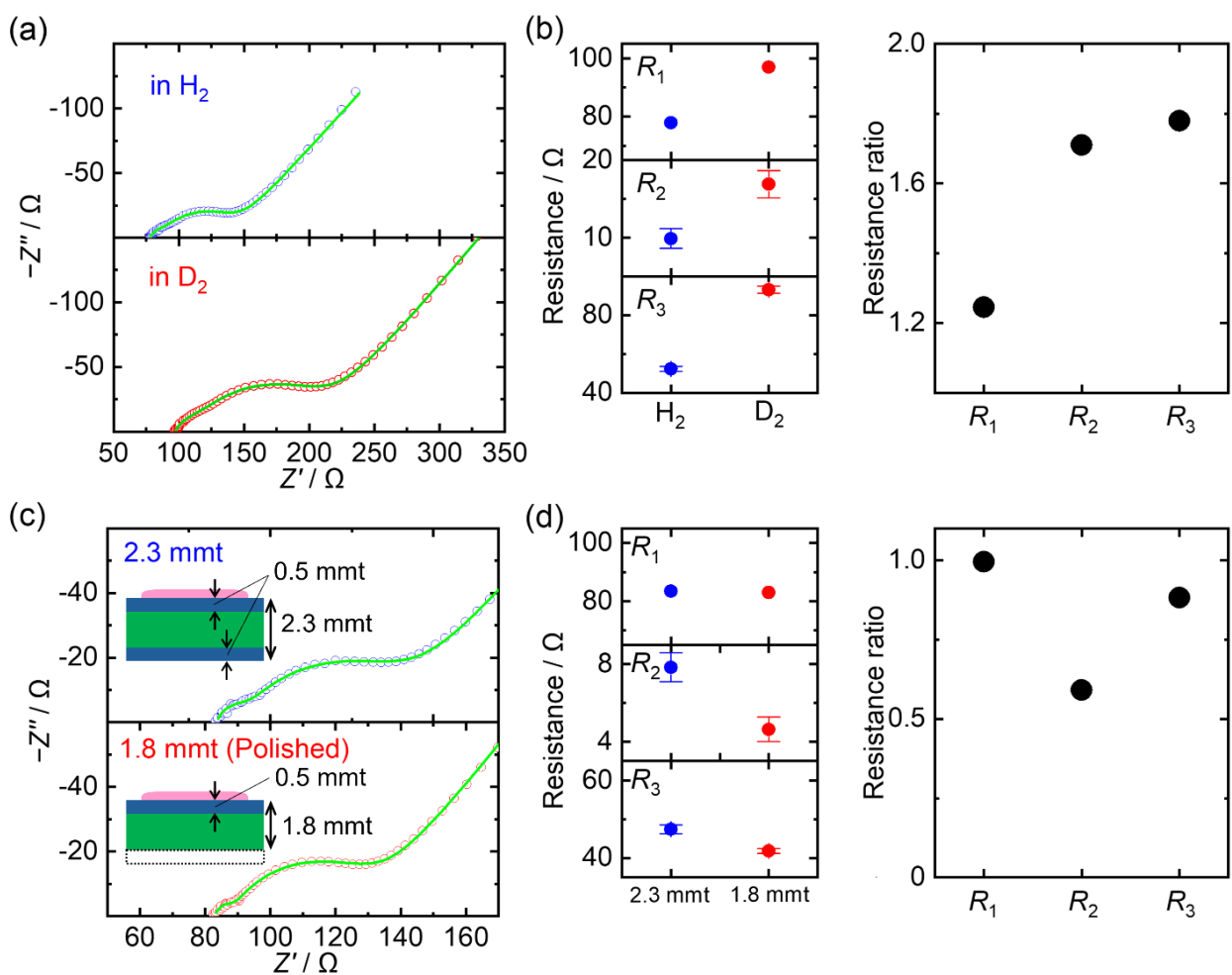

**Figure 4.** (a) Nyquist plots of the impedance spectra measured under H$_2$ (top) and D$_2$ (bottom) atmospheres at 250 °C. (b, left) Fitted resistance values of $R_1$, $R_2$, and $R_3$ under H$_2$ and D$_2$. (b, right)

Corresponding resistance ratios ($R(D_2) / R(H_2)$) representing isotope effects for each component. (c) Nyquist plots of impedance spectra measured under $H_2$ at 250 °C for samples with different thicknesses (2.3 mm and 1.8 mm, polished). (d, left) Fitted resistance values of $R_1$, $R_2$, and $R_3$ for 2.3-mm- and 1.8-mm-thick samples. (d, right) Corresponding resistance ratios ($R(1.8\ \text{mmt}) / R(2.3\ \text{mmt})$) for each component.

Figure 5(a) displays the Arrhenius plot of the proton conductivity ($\sigma_H$) of $H_xWO_3$ measured in a hydrogen atmosphere. The conductivity was calculated for the hydrogen-rich $H_{0.24}WO_3$ region (thickness: 1 mm) from the $R_2$ component. The corresponding electronic conductivity ($\sigma_e$), determined by the DC four-probe method (Figure S8(a)), is also plotted. Although the proton transport number in the measured temperature range is relatively low ($<1 \times 10^{-3}$), $\sigma_H$ reaches $1 \times 10^{-1}$ S cm$^{-1}$ at 275 °C. This value significantly exceeds that of perovskite-type proton-conducting electrolytes in the intermediate-temperature range[52–54], indicating that $H_xWO_3$ holds promise as an electrode material when combined with such electrolytes.

Figure 5(b) presents the temperature dependence of the proton diffusion coefficient ($D_H$) in $H_{0.24}WO_3$ calculated from proton conductivity and proton carrier density using the Nernst–Einstein relation:

$$D_H = \frac{k_B T \sigma_H}{e^2 n_H} \quad (3)$$

where $k_B$ is the Boltzmann constant, and $T$ is the absolute temperature, $e$ is the elementary charge. The value of $D_H$ extrapolated to lower temperatures agrees well with that reported in an earlier work[30], indicating that the excellent proton transport properties are retained over a wide temperature range.

Herein, we compare the proton diffusion behavior of $H_xWO_3$ with that of rutile-type $H_xTiO_2$, which also incorporates protons via hydrogen dissolution (Equation (1))[55,56]. Like $WO_3$, $TiO_2$ undergoes a transition from a $d^0$ to a $d^1$ electronic configuration upon hydrogen incorporation, leading to the formation of $Ti^{3+}$ species. However, the amount of hydrogen incorporated into $TiO_2$ is significantly smaller than that in $WO_3$, typically resulting in proton densities on the order of $10^{18}$–$10^{19}$ cm$^{-3}$ corresponding to a hydrogen composition of $x \sim 0.0001$[36]. In both $H_xWO_3$ and $H_xTiO_2$, equal amounts of protons and electrons are introduced by hydrogen dissolution, leading to coupled proton–electron transport, referred to as ambipolar diffusion[57]. Notably, the $D_H$ of $H_{0.24}WO_3$ is 100–1000 times greater than that of $H_{\sim 0.0001}TiO_2$ along its one-dimensional $c$-axis diffusion pathways[36,58]. In $H_xTiO_2$, the electrons tend to localize at Ti sites, forming small polarons ($Ti_{Ti}$') with low mobility[59,60]. The negative charge of these polarons electrostatically attracts protons and acts as proton traps, thereby limiting the proton mobility. In contrast, in $H_xWO_3$, the electrons introduced via hydrogen dissolution are likely to form polarons with a relatively larger polaron radius, as suggested by the small activation energies ($E_a$) for electronic conduction (Figure S8(b))[61–63]. These highly mobile and delocalized electrons do not produce strong localized electrostatic fields that immobilize protons, accounting for the larger $D_H$ of $H_{0.24}WO_3$ than that of $H_{\sim 0.0001}TiO_2$.

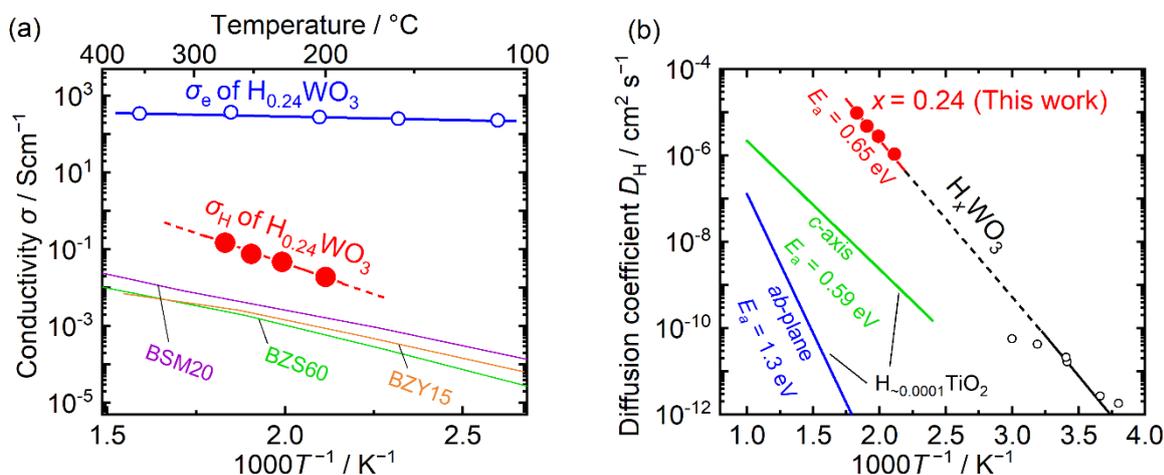

**Figure 5.** (a) Arrhenius plot of conductivity. The proton conductivity ($\sigma_H$) of $H_{0.24}WO_3$ is shown as red closed circles, and the electronic conductivity ($\sigma_e$) as blue open circles. For comparison, $\sigma_H$ values of proton-conducting oxide electrolytes are also plotted: $BaSc_{0.8}Mo_{0.2}O_{2.8}$ (BSM20; purple line)[52], $BaZr_{0.4}Sc_{0.6}O_{3-\delta}$ (BZS60; green line)[53], and $BaZr_{0.85}Y_{0.15}O_{3-\delta}$ (BZY15; orange line)[54]. (b) Arrhenius plot of the proton diffusion coefficient ($D_H$) and activation energy ($E_a$). The red circles represent the values calculated from the $\sigma_H$ of $H_{0.24}WO_3$ measured in this study. Black open circles are the values for crystalline $H_xWO_3$ thin film ($x \sim 0.1$), as reported by Randin et al[30]. For comparison, $D_H$ along the ab-plane and c-axis of $H_{\sim 0.0001}TiO_2$, reported by Johnson et al., are shown as blue and green lines, respectively[36].

Interestingly, $H_{0.24}WO_3$ exhibits $E_a$ for proton diffusion that is comparable to or even higher than that of $H_{\sim 0.0001}TiO_2$ along the c-axis, despite the significantly higher diffusion coefficient. The temperature dependence of diffusion coefficient can be determined using the Arrhenius equation:

$$D_H = D_0 \exp\left(-\frac{E_a}{k_B T}\right) \quad (4)$$

where $D_0$ is the pre-exponential factor. The large $D_H$ in $H_{0.24}WO_3$ associated with the formation of a large polaron can be attributed to a larger $D_0$ compared to that of $H_{\sim 0.0001}TiO_2$, as shown in Table 1. From the random-walk theory, $D_0$ is described as follows:

$$D_0 = \frac{a^2}{6} v_0 \exp\left(\frac{\Delta S_a}{k_B}\right) \quad (5)$$

where $a$ is the jump distance, $v_0$ is the attempt frequency (i.e., the number of attempted jumps per unit time), and $\Delta S_a$ is the activation entropy. For proton conducting oxides, $a$ can be taken as the distance between the neighboring oxide ions where proton transfer occurs, and $v_0$ can be related to be the frequencies of M–O vibrational modes, such as O–M–O bending or M–O stretching modes[64,65]. Thus, for $H_xWO_3$ and $H_xTiO_2$, $a$ and $v_0$ are on the order of 2.7 Å and $10^{13}$ s$^{-1}$, respectively, allowing a rough estimation of $\Delta S_a$ from the experimentally obtained $D_0$ values. As seen in Table 1, the estimated $\Delta S_a$ is larger for $H_{0.24}WO_3$ than that for $H_{\sim 0.0001}TiO_2$. While this difference may partly result from a greater number of accessible diffusion pathways of protons in $H_{0.24}WO_3$, it may also reflect a higher statistical multiplicity of microstates associated with the absorption of multiple phonons required to overcome a migration barrier[66]. A detailed understanding of this behavior remains a subject for future investigation.

**Table 1.** Comparison of diffusion pre-exponential factors and estimated activation entropies for $H_{0.24}WO_3$ and $H_{\sim0.0001}TiO_2$.

|  | $H_{0.24}WO_3$ | $H_{\sim0.0001}TiO_2$ ($c$-axis) |
|---|---|---|
| $D_0$ / cm$^2$ s$^{-1}$ | 10 | 0.002 |
| $\Delta S_a$ / meV K$^{-1}$ | 0.8 | 0.04 |

CONCLUSIONS

This study reported on the proton conductivity of hydrogen tungsten bronze ($H_xWO_3$) in the intermediate-temperature range using an electron-blocking method with a phosphate glass electrolyte. The sample exhibited a spatially inhomogeneous hydrogen distribution, consisting of a hydrogen-rich surface region ($x = 0.24$) and a hydrogen-poor core ($x = 0.0048$). The limited penetration depth of hydrogen (~500 μm) even after prolonged annealing suggests that structural phase boundaries may hinder uniform hydrogen incorporation. The impedance component corresponding to proton conduction in $H_{0.24}WO_3$ was separated from the overall impedance based on comparative measurements obtained using isotope substitution and pellets with different thicknesses. The resulting proton conductivity of $H_{0.24}WO_3$ reached $10^{-1}$ S cm$^{-1}$ at 275 °C under a $H_2$ atmosphere. This value significantly exceeds that of perovskite-type proton-conducting electrolytes in the intermediate-temperature range, which suggests that $H_xWO_3$ is a promising electrode material for use in combination with these electrolytes. A comparison with rutile-type $H_{\sim0.0001}TiO_2$ revealed that $H_{0.24}WO_3$ possesses proton diffusion coefficients that are 100–1000 times larger, indicating the formation of large polarons of electrons. The delocalized electrons do not electrostatically trap protons at specific positions, thereby allowing efficient proton migration. Furthermore, $H_{0.24}WO_3$ exhibits a larger $D_0$ than that of $H_{\sim0.0001}TiO_2$, which is attributed to a greater $\Delta S_a$. This phenomenon likely reflects the presence of a wide variety of accessible proton migration pathways or reduced vibrational constraints in the activated state in $H_{0.24}WO_3$, contributing to its high proton diffusivity despite a comparable or higher $E_a$. These results highlight the critical role of entropic contributions in the proton diffusion process of $H_xWO_3$, offering a new perspective for designing proton-conducting materials with high proton conductivity.

ASSOCIATED CONTENT

**Supporting Information**.

Microstructure of sintered $H_xWO_3$, hydrogen distribution in the $H_xWO_3$ measured by TOF-SIMS, Quantification of hydrogen in the sintered $H_xWO_3$ by TDS, Growth behavior of the hydrogen-rich region by hydrogen annealing, Details of the electron-blocking measurement, and electronic conductivity of $H_xWO_3$.

AUTHOR INFORMATION

Corresponding Author

　**Tomoyuki Yamasaki** – *Institute of Multidisciplinary Research for Advanced Materials (IMRAM), Tohoku University, 2-1-1 Katahira, Aoba-ku, Sendai 980-8577, Japan;* ORCID *orcid.org/*0000-0002-6982-7538; Email: tomo.yamasaki@tohoku.ac.jp

**Takahisa Omata** – *Institute of Multidisciplinary Research for Advanced Materials (IMRAM), Tohoku University, 2-1-1 Katahira, Aoba-ku, Sendai 980-8577, Japan; ORCID orcid.org/*0000-0002-6034-4935; Email: takahisa.omata.c2@tohoku.ac.jp

Author Contributions

Rantaro Matsuo: Investigation, Visualization. Tomoyuki Yamasaki: Investigation, Visualization, Funding acquisition, Writing – original draft. Takahisa Omata: Conceptualization, Funding acquisition, Supervision, Writing – review & editing.

Notes

The authors declare no competing financial interest.

ACKNOWLEDGMENT

This work was supported in part by JSPS KAKENHI Grant Numbers 21H04607 and 24K17762. This work was partly performed under the Cooperative Research Program of the "Network Joint Research Center for Materials and Devices" (No. 20241114) and "Dynamic Alliance for Open Innovation Bridging Human, Environment, and Materials".

ABBREVIATIONS

MIEC, mixed ionic–electronic conductor; QMASS, quadrupole mass spectrometer; SPS, spark plasma sintering; TOF-SIMS, time-of-flight secondary ion mass spectrometry; TDS, thermal desorption spectrometry; XRD, X-ray diffraction.

# High Proton Conductivity of H$_x$WO$_3$ at Intermediate Temperatures: Unlocking Its Application as a Mixed Ionic–Electronic Conductor

Rantaro Matsuo[1], Tomoyuki Yamasaki[1,*], and Takahisa Omata[1,*]

[1] *Institute of Multidisciplinary Research for Advanced Materials (IMRAM), Tohoku University, 2-1-1 Katahira, Aoba-ku, Sendai 980-8577, Japan.*

## S1. Microstructure of sintered H$_x$WO$_3$

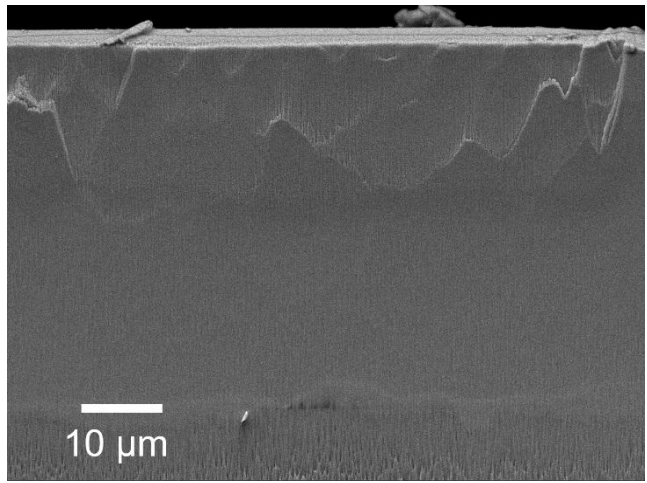

**Figure S1.** Cross-sectional SEM image of H$_x$WO$_3$. The cross-sectional surface was prepared using a cross-section polisher after cutting the hydrogenated WO$_3$ pellet. Needle-like features were observed on the cross-sectional surface as a result of ion-beam polishing. Grain boundaries were too fine to be resolved, suggesting a compact and homogeneous microstructure composed of very fine crystallites.

## S2. Hydrogen distribution in the sintered $H_xWO_3$ measured by TOF-SIMS

### S2.1. Background measurement of hydrogen in $WO_3$

To evaluate the background hydrogen signal in the TOF-SIMS analysis and determine the necessary sputtering time required to eliminate surface contamination, depth profile measurements were conducted on a reference $WO_3$ sintered pellet (annealed in flowing oxygen at 800 °C for 10 h) as well as on the $H_xWO_3$ pellet. The TOF-SIMS depth profiles of the secondary ions were acquired in the negative polarity mode using a $Bi^+$ primary ion beam (25 kV, 1.0 pA) and a $Cs^+$ sputter ion beam (2 kV, 145 nA). Figure S2(a) shows the depth profiles of $H^-$ and $OH^-$ signals, normalized to the $W^-$ signal for reference $WO_3$ and $H_xWO_3$ pellets. In the reference $WO_3$ sample, the signals gradually decayed toward the interior of the pellet and became nearly constant after approximately 60 min of sputtering, indicating the elimination of the contribution from surface-adsorbed species. In contrast, for the $H_xWO_3$ sample, the intensities of $H^-$ and $OH^-$ signals remained nearly constant throughout the sputtering process. Even after sputtering beyond the depth at which surface contributions were no longer observed in the reference $WO_3$ sample, the signal intensities in $H_xWO_3$ remained approximately an order of magnitude higher than those in the reference. These results clearly demonstrate that hydrogen in the $H_xWO_3$ sample can be reliably detected above the background level observed in the reference $WO_3$ sample.

### S.2.2. SIMS area imaging of the cross-section of $H_xWO_3$

To obtain SIMS imaging data free from surface contamination, a thin surface layer was sputtered prior to acquisition, and imaging was initiated from the depth at which the signal intensities of $H^-$ and $OH^-$ became stable. The imaging area was selected 300 μm inward from the edge of the pellet. A $500 \times 500$ μm$^2$ region was imaged to visualize the lateral distribution of hydrogen. Figure S2(b) displays lateral SIMS images of $H^-$, $OH^-$, and $W^-$. In these images, the left edge corresponds to a depth of 300 μm from the surface, and the right edge corresponds to approximately 800 μm. To convert the SIMS images into a concentration profile along the thickness direction of the sintered pellet (Figure 1 (b)), the signal intensities were integrated along the Y-direction.

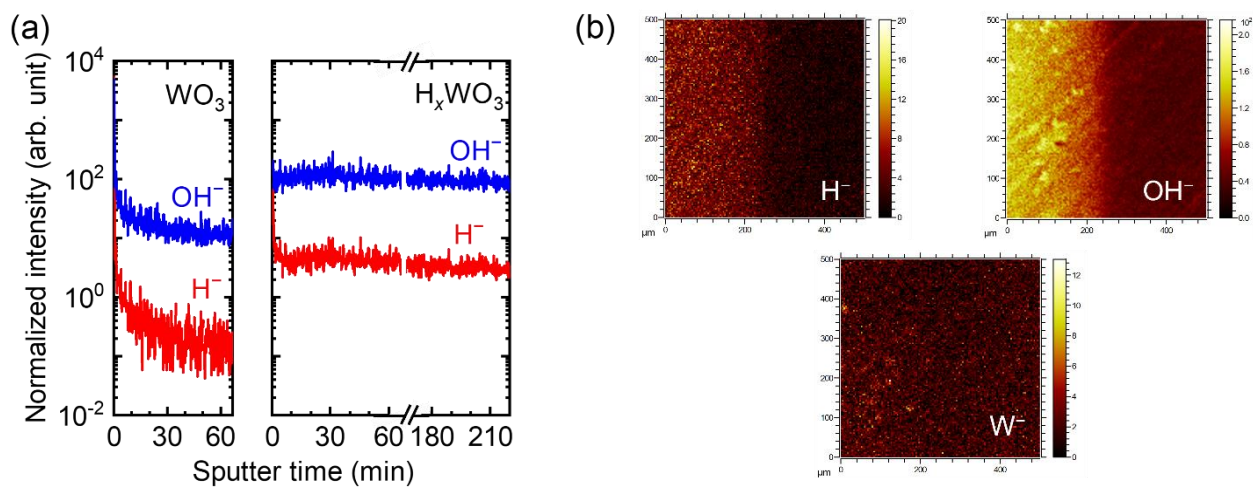

**Figure S2.** (a) Depth profiles of H⁻ and OH⁻ signals normalized to the W⁻ signal for reference $WO_3$ and $H_xWO_3$ pellets. (b) Lateral SIMS images of H⁻, OH⁻, and W⁻ in the vicinity of the phase boundary in $H_xWO_3$.

## S3. Quantification of hydrogen in sintered $H_xWO_3$ by TDS

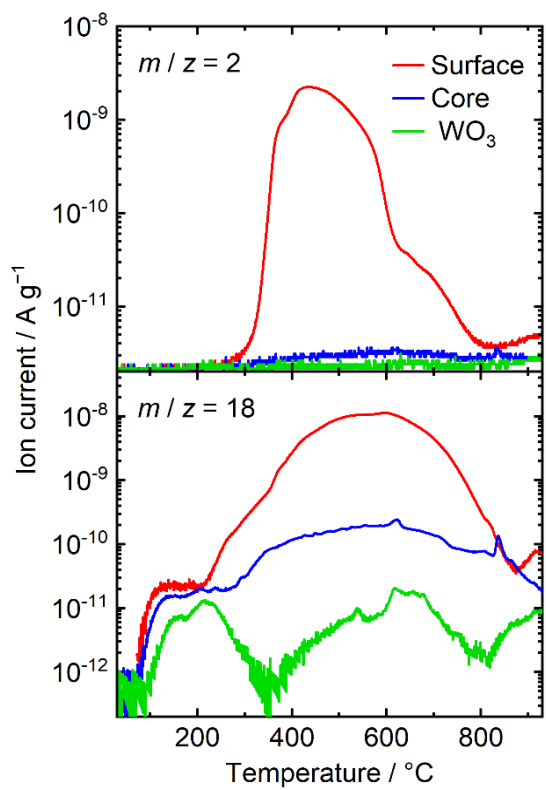

**Figure S3.** TDS curves of $m/z = 2$ ($H_2$) and 18 ($H_2O$) for the hydrogen-rich surface (red) and hydrogen-poor core (blue) regions of $H_xWO_3$ along with $WO_3$ pellet annealed in oxygen at 800 °C as a reference (green). The release of hydrogen from $H_xWO_3$ in the form of not only $H_2$ but also $H_2O$ is likely due to heating under vacuum conditions.

**Table S1.** Summary of the quantification of hydrogen released as $H_2$ and $H_2O$.

|  | Surface | Core |
| --- | --- | --- |
| H released as $H_2$ / $cm^{-3}$ | $1.3 \times 10^{21}$ | $3.2 \times 10^{17}$ |
| H released as $H_2O$ / $cm^{-3}$ | $3.2 \times 10^{21}$ | $8.9 \times 10^{19}$ |
| Total H / $cm^{-3}$ | $4.5 \times 10^{21}$ | $8.9 \times 10^{19}$ |
| $x$ in $H_xWO_3$ | 0.24 | 0.0048 |

**S4. Growth behavior of the hydrogen-rich region by hydrogen annealing**

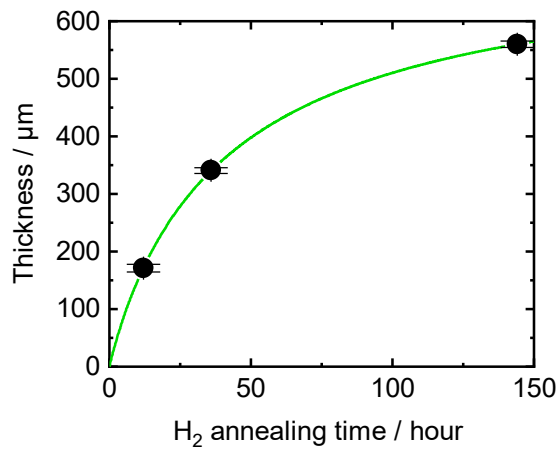

**Figure S4.** Thickness variation of the hydrogen-rich region as a function of annealing time. The green curve represents a fitting result. The sintered $WO_3$ samples were annealed in a hydrogen atmosphere at 300 °C for varying durations. The thickness of the hydrogen-rich region, measured from the Pd-deposited surface, gradually increases with annealing time.

## S5. Supporting details of the electron-blocking measurement

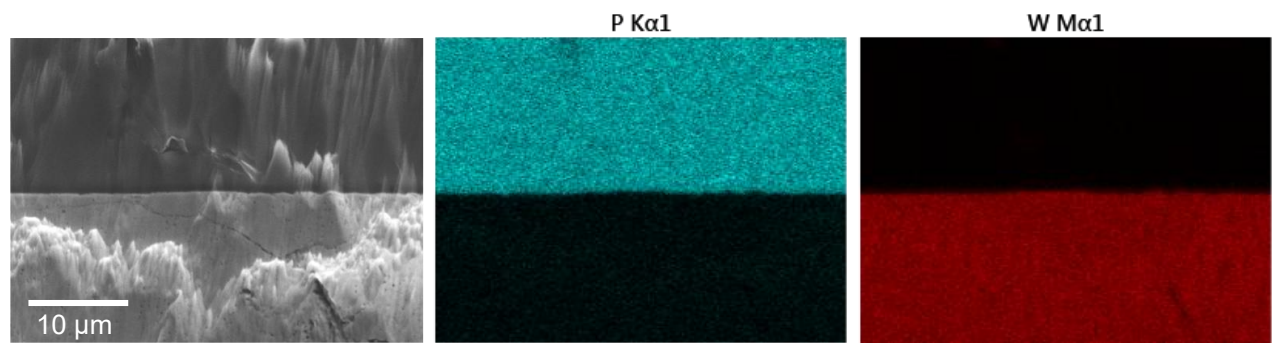

**Figure S5.** Cross-sectional SEM image of the electron-blocking cell and the corresponding elemental maps for P Kα and W Mα obtained by energy-dispersive X-ray spectroscopy (EDX). The elemental distributions confirm the presence of the phosphate glass electrolyte and the $H_xWO_3$ layer. No delamination or gap is observed at the interface, indicating that the phosphate glass and $H_xWO_3$ are well adhered.

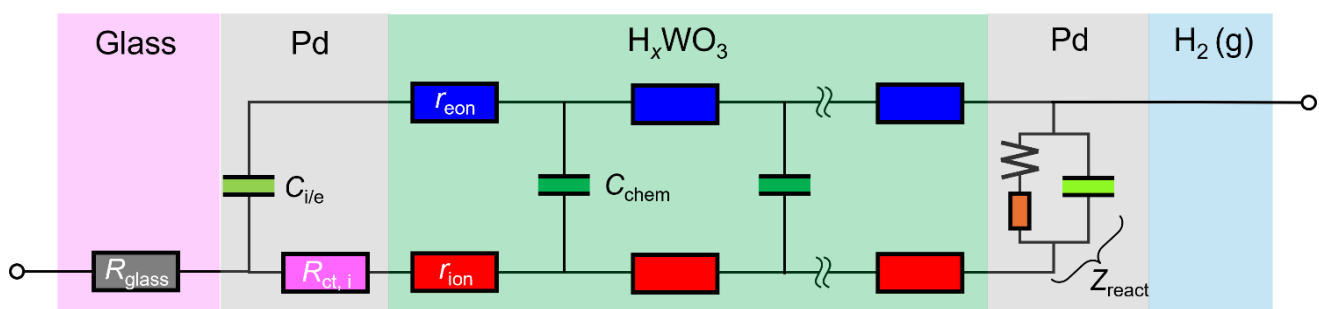

**Figure S6.** Illustration of the equivalent circuit for the asymmetric electron-blocking cell based on the transmission line model. The model includes ionic and electronic conduction pathways in the $H_xWO_3$ pellet, as well as charge transfer processes at the glass blocking electrode interface and reactions at the gas–solid interface. The effects of hydrogen distribution inhomogeneity or the presence of multiple phases in $H_xWO_3$ are not considered.

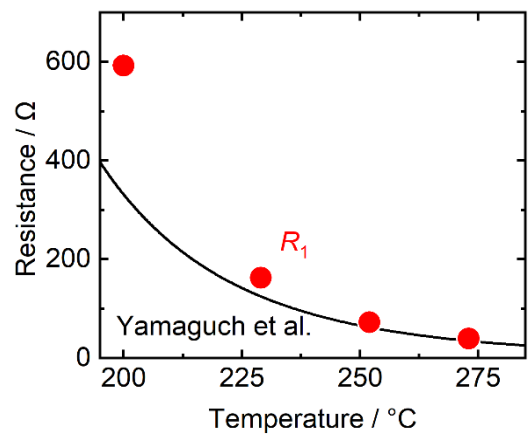

**Figure S7.** Temperature dependence of $R_1$ compared with the resistance calculated for the glass with the same geometry using the reported conductivity data[S1].

## S6. Electronic conductivity of H$_x$WO$_3$

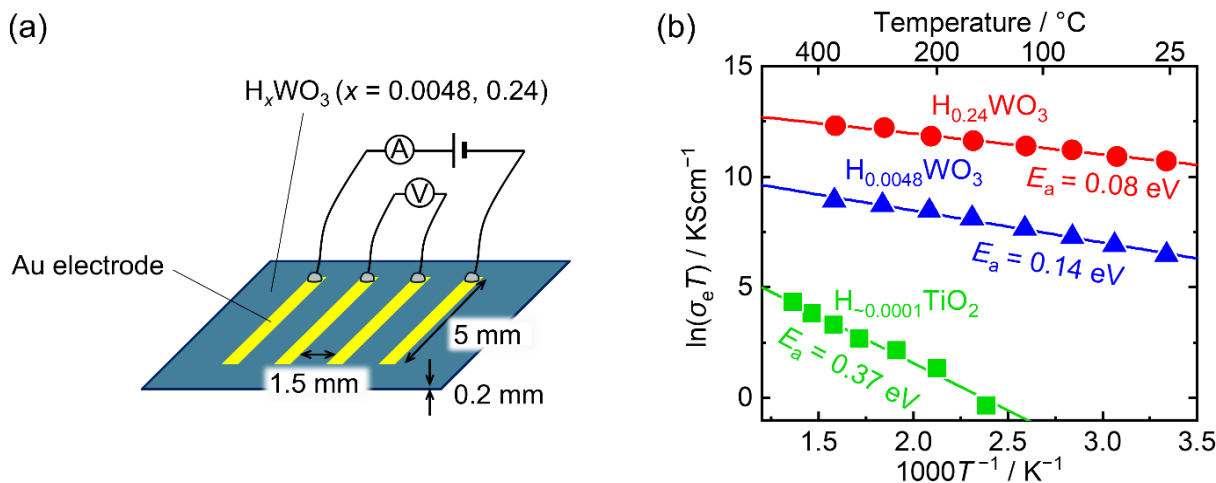

**Figure S8.** (a) Illustration of the planar sample for electronic conductivity measurements using the four-probe method. (b) Arrhenius plots of the electronic conductivity for H$_{0.24}$WO$_3$ (red circle), H$_{0.0048}$WO$_3$ (blue triangle), and H$_{\sim0.0001}$TiO$_2$ (green square) with the corresponding activation energies[S2]. The electronic conductivities of H$_{0.24}$WO$_3$ and H$_{0.0048}$WO$_3$ were measured under Ar to prevent changes in hydrogen content, while that of H$_{\sim0.0001}$TiO$_2$ was measured under H$_2$ atmosphere.